\documentclass[a4paper,11pt]{article}
\pdfoutput=1 

\usepackage{jheppub} 

\usepackage[T1]{fontenc} 
\usepackage{slashed}
\usepackage{physics}
\usepackage{mathbbol} 
\usepackage{bm}
\usepackage{xcolor}

\title{\boldmath Can Bell inequalities be tested via scattering cross-section at colliders ?}

\author[a,b]{Song Li,}
\author[a,b]{Wei Shen,}
\author[a,c]{Jin Min Yang}

\affiliation[a]{CAS Key Laboratory of Theoretical Physics, Institute of Theoretical Physics, Chinese Academy of Sciences, Beijing 100190, China}
\affiliation[b]{School of Physics, University of Chinese Academy of Sciences,  Beijing 100049, China}
\affiliation[c]{School  of Physics, Henan Normal University, Xinxiang 453007, P. R. China}

\emailAdd{lisong@itp.ac.cn}
\emailAdd{shenwei@itp.ac.cn}
\emailAdd{jmyang@itp.ac.cn}

\abstract{
In current studies for testing Bell inequalities at colliders, the reconstruction of spin correlations from scattering cross-sections relies on the bilinear form of the spin correlations, but not all local hidden variable models (LHVMs) have such a property. To demonstrate that a general LHVM cannot be rule out via scattering cross-section data, we propose a specific LHVM, which can exactly duplicate the same scattering cross-section for particle production and decay as the standard quantum theory, making it indistinguishable at colliders in principle. Despite of this, we find that reconstructing spin correlations through scattering cross-sections can still exclude a broad class of LHVMs, e.g., those models employing classical spin correlations as a surrogate for quantum spin correlations.           
}

\begin{document} 
\maketitle
\flushbottom

\section{Introduction}
\label{sec:intro}

Bell theorem states that constraints on the correlation of non-commuting observables in a local hidden variable model (LHVM) are incompatible with quantum mechanics~\cite{Bell:1964kc}, and Bell inequalities have been tested in various non-relativistic entangled quantum states~\cite{freedman1972experimental,aspect1982experimental,big2018challenging}. Recently, some studies suggest that high-energy colliders can also provide an ideal platform for studying quantum information in the relativistic region~\cite{Giacomini_2019, Afik_2022}. Specifically, it has been proposed that Bell inequalities can be tested via the scattering cross-section of particle pair production and decay at colliders~\cite{Qian:2020ini,Fabbrichesi:2021npl,Han:2023fci,Ehataht:2023zzt}. We will briefly review the methods currently being employed by researchers to explore this intriguing topic in the following.

To extract the spin correlation information of the collision products from the data of the scattering cross section, we need to start with the production spin density matrix and the decay spin density matrix. For example, consider the process where particles $\mathcal{X}$ and $\mathcal{Y}$ collide to produce particles $\mathcal{A}$ and $\mathcal{B}$, which then decay into $a_1+a_2+a_3$ and $b_1+b_2+b_3$ respectively in an extremely short time. In our subsequent discussions, the number of decay products is not crucial and will not affect the final conclusion. Under the narrow width approximation, we have the following differential scattering cross section:
\begin{eqnarray}
&&\dv{\sigma}{\Pi_{a_{1\sim 3}}\dd\Pi_{b_{1\sim 3}}}\,(\mathcal{XY}\to\mathcal{AB}\to(a_1a_2a_3)(b_1b_2b_3)) \nonumber \\
\propto && \sum_{s,r,\bar{s},\bar{r}}\overline{\sum_{\rm in}}\mathcal{M}_{\mathcal{XY}\to\mathcal{A}^s\mathcal{B}^{\bar{s}}}\mathcal{M}^*_{\mathcal{XY}\to\mathcal{A}^r\mathcal{B}^{\bar{r}}}
\mathcal{M}_{\mathcal{A}^r\to a_1a_2a_3}\mathcal{M}^*_{\mathcal{A}^s\to a_1a_2a_3}
\mathcal{M}_{\mathcal{B}^{\bar{r}}\to b_1b_2b_3}\mathcal{M}^*_{\mathcal{B}^{\bar{s}}\to b_1b_2b_3}\nonumber \\
= &&  \sum_{s,r,\bar{s},\bar{r}}R_{sr,\bar{s}\bar{r}}\Gamma^\mathcal{A}_{rs}\Gamma^\mathcal{B}_{\bar{r}\bar{s}}\,,
\end{eqnarray}
where $R_{sr,\bar{s}\bar{r}}=\overline{\sum}_{\rm in}\mathcal{M}_{\mathcal{XY}\to\mathcal{A}^s\mathcal{B}^{\bar{s}}}\mathcal{M}^*_{\mathcal{XY}\to\mathcal{A}^r\mathcal{B}^{\bar{r}}}$ is the \textit{production spin density matrix}, with $s,r$ ($\bar{s},\bar{r}$) being the spin indices of particle $\mathcal{A}$ ($\mathcal{B}$), whose corresponding spin directions can be specified by convention. $\Gamma^\mathcal{A}_{rs}$ and $\Gamma^\mathcal{B}_{\bar{r}\bar{s}}$ are the \textit{decay spin density matrices}~\cite{Bernreuther:2004jv,Fabbrichesi:2022ovb} for $\mathcal{A}$ and $\mathcal{B}$, respectively, given by 
\begin{eqnarray}
    \Gamma^\mathcal{A}_{rs} &=& \mathcal{M}_{\mathcal{A}^r\to a_1a_2a_3}\mathcal{M}^*_{\mathcal{A}^s\to a_1a_2a_3}\,, \\
    \Gamma^\mathcal{B}_{\bar{r}\bar{s}} &=& \mathcal{M}_{\mathcal{B}^{\bar{r}}\to b_1b_2b_3}\mathcal{M}^*_{\mathcal{B}^{\bar{s}}\to b_1b_2b_3}\,.
\end{eqnarray}
 For simplicity, we assume both $\mathcal{A}$ and $\mathcal{B}$ are spin-1/2.

We can express $R_{sr,\bar{s}\bar{r}}$ as
\begin{equation}
R_{sr,\bar{s}\bar{r}} \propto \mel*{\mathcal{A}^r\mathcal{B}^{\bar{r}}}{\left(\overline{\sum_{\rm in}}\ketbra{\mathrm{in}}\right)}{\mathcal{A}^s\mathcal{B}^{\bar{s}}}\,.
\end{equation}
From this expression, it is evident that $R_{sr,\bar{s}\bar{r}}$ indeed has the physical meaning of a density matrix, containing all the spin correlation information of $\mathcal{A}$ and $\mathcal{B}$. If the information of $R_{sr,\bar{s}\bar{r}}$ can be extracted from the scattering cross section data, it can be used for the study of Bell inequalities.

It is apparent that $\Gamma^{\mathcal{A}}_{rs}$, with $r,s$ as its row and column indices, can form a $2\times2$ Hermitian matrix, thus it can be expanded using the basis $\{\mathbb{1},\sigma_1,\sigma_2,\sigma_3\}$. In fact, if we choose the rest frame of particle $\mathcal{A}$, and integrate out all other parts of the decay final state phase space while keeping the direction angle $\mathbf{\Omega}^\mathcal{A}$ of one of the decay products fixed, then by utilizing the properties of the representation of the rotation group, we can obtain
\begin{equation}
\frac{\tilde{\Gamma}^\mathcal{A}_{rs}(\mathbf{\Omega}^\mathcal{A})}{\Gamma^\mathcal{A}}=\frac12\left(\delta_{rs}+\kappa^\mathcal{A}\Omega^\mathcal{A}_i\sigma_{i,rs}\right)\,,
\end{equation}
where $\Gamma^\mathcal{A}$ is the decay width of $\mathcal{A}$, and $\kappa^\mathcal{A}$ contains the polarization information of $\mathcal{A}$, which can be obtained from the standard process of calculating scattering amplitudes. Here we used the Einstein summation convention over the spatial index $i$, summing from $1$ to $3$. $\Omega_i^\mathcal{A}$ are the components of $\mathbf{\Omega}^\mathcal{A}$ in the rest frame of $\mathcal{A}$, decomposed into three orthogonal unit vectors as the basis. The choice of the three basis vectors is arbitrary, however, we can make a deliberate selection so that the three basis vectors in different emission directions in the experimental reference frame of $\mathcal{X}\mathcal{Y}\to\mathcal{A}\mathcal{B}$ can be aligned by rotation matrices.

The density operator of a two-qubit system can always be parameterized as
\begin{equation}
\rho=\frac14\left(\mathbb{1}_4+B_i^\mathcal{A}(\sigma_i\otimes\mathbb{1})+B_i^\mathcal{B}(\mathbb{1}\otimes\sigma_i)+C_{ij}(\sigma_{i}\otimes\sigma_j)\right)\,.
\end{equation}
Thus, we can obtain~\cite{Bernreuther:2004jv,Fabbrichesi:2022ovb}
\begin{equation}\label{eq:EqforCij}
	\frac{1}{\sigma}\frac{\dd{}^4\sigma}{\dd{}^2\Omega^\mathcal{A}\dd{}^2\Omega^\mathcal{B}} =\frac{1}{(4\pi)^2}\big(1+\kappa^\mathcal{A}B_i^\mathcal{A}\Omega_i^\mathcal{A}+\kappa^\mathcal{B}B_i^\mathcal{B}\Omega_i^\mathcal{B} +\kappa^\mathcal{A}\kappa^\mathcal{B}\Omega_i^\mathcal{A}C_{ij}\Omega_j^\mathcal{B}\big)\,.
\end{equation}
In principle, we could extract the coefficients of the density operator based on the collider event data with this result, but we need not delve into the numerical analysis process here.

If we measure the spin of particle $\mathcal{A}$ in the direction $\mathbf{n}^a$ and the spin of particle $\mathcal{B}$ in the direction $\mathbf{n}^b$, then the spin correlation will be $P(\mathbf{n}^a,\mathbf{n}^b)=n^a_iC_{ij}n_j^b$. However, we cannot directly measure spin at a collider; instead, we rely on Eq.~\eqref{eq:EqforCij} to obtain the spin correlation coefficients $C_{ij}$. The current approaches claim that Bell inequalities can be reduced to constraints solely on the matrix $C$. Among these constraints, a typical one is that the sum of the two largest eigenvalues of the matrix $C^{\rm T}C$ should not exceed $1$~\cite{horodecki1995violating}, which has been demonstrated to be violated at a confidence level of 98\% for top quark pairs at the Large Hadron Collider~\cite{Fabbrichesi:2021npl}. 

While the above collider tests for Bell inequalities seem promising, some of these approaches need to be further considered. These approaches are premised on standard quantum theory, so using them to explore the violation of Bell inequalities to rule out a general LHVM would be sort of circular reasoning. One obvious point is that a spin correlation in an LHVM is not necessarily bilinear, and an LHVM describing Bell states is inherently incompatible with spin correlation bilinearity, which will be analyzed in detail in this work. Although it can be argued that testing the violation of Bell inequalities within the framework of quantum theory is just one way to explore the quantum correlations of spin in produced particles and these efforts can serve as a theoretical guidance for future direct tests~\cite{Han:2023fci}, our study in this work is focusing on whether such production and decay cross-sections of non-zero spin particles at colliders can be used to test Bell inequalities to rule out a general LHVM. Early in 1992, an LHVM was proposed to describe the scattering distribution of final decay products and showed that it may not be possible to test Bell inequalities at a collider with available experiment techniques~\cite{Dreiner:1992gt,Abel:1992kz}, whereas so far there has been no such a scrutinization on the recent collider approaches involving the reconstruction of the production density matrix of intermediate particles.

In this work we suppose there is a sufficiently powerful collider detector, which can not only measure the scattering cross-sections at any intermediate stage of the reaction chain but also observe the angular distributions of the decay products after decoherence occurs (we assume that it cannot detect spin). A collider detector of this kind does not exist with current technology and is used here merely as a starting point for our general argument. Such a detector would provide all the spin-summed scattering cross-section information. We will provide a general local hidden-variable description of the production and decay of particles, demonstrating that even with the aid of this powerful detector, we are still unable to carry out a Bell test for general LHVMs. Finally, we will discuss which LHVMs can be ruled out by the methods currently employed by researchers. We find that if an LHVM attempts to interpret the quantum correlations of spin as classical correlations and inherit the results of standard quantum theory regarding particle decay, then it can be excluded by the current methods.

It is well-known that testing Bell's inequalities using colliders has several loopholes, such as the locality loophole and the fair sampling loophole (or the causality loophole and the selection loophole)~\cite{Severi:2021cnj,Han:2023fci}. However, these loopholes are mostly due to technical reasons rather than theoretical difficulties and can be avoided through specific methods. The loophole we point out here, however, stems entirely from the local hidden variable theory itself. As long as the spins in different directions are not measured, no matter how colliders are improved, the test of Bell's inequalities cannot be completed.

\section{Bilinearity of Spin Correlations}
\label{sec:BilinearSpinCorr}

In the current methods, obtaining $C_{ij}$ through scattering cross-sections actually assumes the property of bilinearity in spin correlations. This is because we rely on the relationship $P(\mathbf{n}^a,\mathbf{n}^b)=n_i^aP(\mathbf{e}^a_i,\mathbf{e}^b_j)n_j^b$ to obtain the spin correlations between any directions. However, we do not need to introduce a standard orthogonal basis for this step. We can directly restore the spin correlations in any direction from the scattering cross-section data without using $C_{ij}$. For this purpose, we expand $\mathbf{n}^a$ into a set of unit bases $\{\mathbf{n}_i^a\}$ (not necessarily orthogonal to each other), and let their dual bases be $\{\tilde{\mathbf{n}}_i^a\}$. Similarly, we can also obtain another set of bases $\{\mathbf{n}_i^b\}$ and their dual bases $\{\tilde{\mathbf{n}}_i^b\}$. We define
\begin{alignat}{2}
\sigma^a_i &= \mathbf{n}^a_i\cdot\boldsymbol{\sigma}\,,\qquad &\sigma^b_i &= \mathbf{n}^b_i\cdot\boldsymbol{\sigma}\,;\\
\tilde\sigma^a_i &= \tilde{\mathbf{n}}^a_i\cdot\boldsymbol{\sigma}\,,\qquad &\tilde\sigma^b_i &= \tilde{\mathbf{n}}^b_i\cdot\boldsymbol{\sigma}\,;\\
\Omega_i^a &= \tilde{\mathbf{n}}^a_i\cdot\boldsymbol{\Omega}^{\mathcal{A}}\,,\qquad &\Omega_i^b &= \tilde{\mathbf{n}}^b_i\cdot\boldsymbol{\Omega}^{\mathcal{B}}\;.
\end{alignat}
Under the inner product of Hermitian matrices $\langle AB\rangle=\frac12\tr(AB)$, $\{\sigma_i^{a(b)}\}$ and $\{\tilde{\sigma}_i^{a(b)}\}$ are dual bases to each other. In the new basis, the density matrix of a two-qubit system can be written as
\begin{equation}\label{eq:DM-NewBasics}
\rho=\frac14\left(\mathbb{1}_4+\tilde{B}_i^\mathcal{A}(\tilde{\sigma}^a_i\otimes\mathbb{1})+\tilde{B}_i^\mathcal{B}(\mathbb{1}\otimes\tilde{\sigma}^b_i)+P(\mathbf{n}^a_i,\mathbf{n}^b_j)(\tilde{\sigma}^a_{i}\otimes\tilde{\sigma}^b_j)\right)\,.
\end{equation}
For the decay spin density matrix, we have
\begin{equation}\label{eq:DSDM-NewBasics}
\begin{aligned}
\frac{\tilde{\Gamma}^\mathcal{A}_{rs}(\mathbf{\Omega}^\mathcal{A})}{\Gamma^\mathcal{A}} &= \frac12\left(\delta_{rs}+\kappa^\mathcal{A}\Omega^\mathcal{A}_i\sigma_{i,rs}\right)\\
&= \frac12\left(\delta_{rs}+\kappa^\mathcal{A}(\mathbf{\Omega}^\mathcal{A}\cdot\tilde{\mathbf{n}}_i^a)\sigma^a_{i,rs}\right)\\
&= \frac12\left(\delta_{rs}+\kappa^\mathcal{A}\Omega^a_i\sigma^a_{i,rs}\right)\,,
\end{aligned}
\end{equation}
where $\Omega_i^a$ are the components of the unit vector $\mathbf{\Omega}^\mathcal{A}$ expanded in the basis vectors $\{\mathbf{n}_i^a\}$. From Eq.~\eqref{eq:DM-NewBasics} and Eq.~\eqref{eq:DSDM-NewBasics}, we can immediately obtain
\begin{equation}\label{eq:EqforPnanb}
\frac{1}{\sigma}\frac{\dd{}^4\sigma}{\dd{}^2\Omega^\mathcal{A}\dd{}^2\Omega^\mathcal{B}}=\frac{1}{(4\pi)^2}\left(1+\kappa^\mathcal{A}\tilde{B}_i^\mathcal{A}\Omega_i^a+\kappa^\mathcal{B}\tilde{B}_i^\mathcal{B}\Omega_i^b+\kappa^\mathcal{A}\kappa^\mathcal{B}\Omega_i^a P(\mathbf{n}^a_i,\mathbf{n}^b_j)\Omega_j^b\right)\,.
\end{equation}
With this result, we can directly restore the needed $P(\mathbf{n}^a,\mathbf{n}^b)$ without relying on the relationship $P(\mathbf{n}^a,\mathbf{n}^b)=n_i^aP(\mathbf{e}^a_i,\mathbf{e}^b_j)n_j^b$.

Unfortunately, Eq.~\eqref{eq:EqforPnanb} does not enable us to circumvent the reliance on the bilinear form. To illustrate, suppose an LHVM satisfies Eq.~\eqref{eq:EqforPnanb} for any unit basis $\{\mathbf{n}_i^a\}$ and $\{\mathbf{n}_i^b\}$, then we have
\begin{equation}
\begin{aligned}
\frac{1}{\sigma}\frac{\dd{}^4\sigma}{\dd{}^2\Omega^\mathcal{A}\dd{}^2\Omega^\mathcal{B}} &= \frac{1}{(4\pi)^2}\left(1+\kappa^\mathcal{A}\tilde{B}_i^\mathcal{A}\Omega_i^a+\kappa^\mathcal{B}\tilde{B}_i^\mathcal{B}\Omega_i^b+\kappa^\mathcal{A}\kappa^\mathcal{B}\Omega_i^a P(\mathbf{n}^a_i,\mathbf{n}^b_j)\Omega_j^b\right)\\
&= \frac{1}{(4\pi)^2}\left(1+\kappa^\mathcal{A}B_i^\mathcal{A}\Omega_i^\mathcal{A}+\kappa^\mathcal{B}B_i^\mathcal{B}\Omega_i^\mathcal{B} +\kappa^\mathcal{A}\kappa^\mathcal{B}\Omega_i^\mathcal{A}P(\mathbf{e}^a_i,\mathbf{e}^b_j)\Omega_j^\mathcal{B}\right)\,.
\end{aligned}
\end{equation}
This equation is valid for any direction $\mathbf{\Omega}^\mathcal{A}$ and $\mathbf{\Omega}^\mathcal{B}$, thus
\begin{equation}
\begin{aligned}
\Omega_i^{a}P(\mathbf{n}^a_i,\mathbf{n}^b_j)\Omega_j^{b} &= \Omega_i^\mathcal{A}P(\mathbf{e}^a_i,\mathbf{e}^b_j)\Omega_j^\mathcal{B} \\
&= \Omega^a_k(\mathbf{e}^a_i\cdot\mathbf{n}_k^a)P(\mathbf{e}^a_i,\mathbf{e}^b_j)(\mathbf{e}^b_j\cdot\mathbf{n}_t^b)\Omega^b_t\\
&= \Omega_k^a n_{k,i}^a P(\mathbf{e}^a_i,\mathbf{e}^b_j) n_{t,j}^b\Omega_t^b\,.
\end{aligned}
\end{equation}
We immediately obtain $P(\mathbf{n}^a_k,\mathbf{n}^b_t)=n_{k,i}^aP(\mathbf{e}^a_i,\mathbf{e}^b_j)n_{t,j}^b$. This shows that as long as Eq.~\eqref{eq:EqforPnanb} holds, the spin correlation of this LHVM must exhibit a bilinear form. However, we want to point out that not all LHVMs describing two qubits have bilinear spin correlations.

Here we simply schematize Bell's originally proposed hidden variable model~\cite{Bell:1964kc} as a counterexample, which does not satisfy $P(\mathbf{n}^a,\mathbf{n}^b)=n_i^aP(\mathbf{e}_i^a,\mathbf{e}_j^b)n_j^b$. Suppose for a spin-$1/2$ particle state $\ket{\mathbf{v}}$ that carries a three-dimensional unit vector $\mathbf{m}$ as hidden variable, the measured value of spin along an arbitrary direction $\mathbf{n}$ is
\begin{equation}\label{eq:BellLHVM}
    \begin{cases}
        +1\,,\qquad\text{if } (\mathbf{v}+\mathbf{m})\cdot\mathbf{n}>0\,;\\
        -1\,,\qquad\text{if } (\mathbf{v}+\mathbf{m})\cdot\mathbf{n}<0\,.
    \end{cases}
\end{equation}
For the case where $(\mathbf{v}+\mathbf{m})\cdot\mathbf{n}=0$, the measure of its parameter space is zero, so it can be ignored. We can artificially make $\mathbf{v}^\mathcal{A}=\mathbf{m}^\mathcal{A}=-\mathbf{v}^\mathcal{B}=-\mathbf{m}^\mathcal{B}=\bm{\lambda}$ to describe the Bell state $(\ket{\uparrow}\ket{\downarrow}-\ket{\downarrow}\ket{\uparrow})/\sqrt{2}$, where $\bm{\lambda}$ is a vector uniformly distributed on a unit sphere. Obviously, for Bell state within this LHVM, the spin correlation does not have the bilinear form:
\begin{equation}\label{eq:ResultofCounterexample}
P(\mathbf{n}^a,\mathbf{n}^b)=-1+\frac{2}{\pi}\arccos(\mathbf{n}^a\cdot\mathbf{n}^b) \,.
\end{equation}

However, there exists a large class of LHVMs that do have bilinear spin correlations. For instance, according to Gleason's theorem~\cite{gleason1975measures}, for a noncontextual LHVM of a two-qubit system, there exist a positive semi-definite Hermitian operator $\rho_\psi$ for each state $\ket{\psi}$ such that for each projection operator $\hat{P}$, the probability of its corresponding measurement outcome is $P(\psi,\hat{P})=\mathrm{tr}(\rho_\psi\hat{P})$. Although $\rho_\psi$ may not be equal to $\ketbra{\psi}$, we can always make the following derivation:
\begin{equation}
\begin{aligned}
P(\mathbf{n}^a,\mathbf{n}^b) &= P(\psi,P^{ab}_{++})+P(\psi,P^{ab}_{--})-P(\psi,P^{ab}_{+-})-P(\psi,P^{ab}_{-+})\\
&= \tr(\rho_\psi P^{ab}_{++})+\tr(\rho_\psi P^{ab}_{--})-\tr(\rho_\psi P^{ab}_{+-})-\tr(\rho_\psi P^{ab}_{-+})\\
&= \tr(\rho_\psi(P^{ab}_{++}+P_{--}^{ab}-P_{+-}^{ab}-P_{-+}^{ab}))\\
&= \tr(\rho_\psi((n^a_i\sigma_i)\otimes(n_j^b\sigma_j))) \\
&= n_i^a \tr(\rho_\psi(\sigma_i\otimes\sigma_j)) n_j^b\\
&= n_i^aP(\mathbf{e}_i^a,\mathbf{e}_j^b)n_j^b\,.
\end{aligned}
\end{equation}
Therefore, the spin correlations of noncontextual LHVMs do indeed possess bilinearity. 

If the spin correlations of an LHVM are in bilinear form, then Bell's inequalities will impose stricter constraints. To illustrate, take
\begin{equation}
    \mathbf{n}^a=\mathbf{e}_i^a\,,\quad \mathbf{n}^{\prime a}=\mathbf{e}^{a}_j\,,\quad \mathbf{n}^b=\frac{1}{\sqrt{2}}(\pm \mathbf{e}_k^b+\mathbf{e}_t^b)\,, \quad \mathbf{n}^{\prime b}=\frac{1}{\sqrt{2}}(\pm \mathbf{e}_k^b-\mathbf{e}_t^b)\,,
\end{equation}
substitute them into the CHSH inequality~\cite{Clauser:1969ny} and use the bilinear property of spin correlations, we can obtain~\cite{Han:2023fci}
\begin{equation} \label{eq:StringentRestrictions4PHVM}
\abs{P(\mathbf{e}^a_i,\mathbf{e}^b_k)\pm P(\mathbf{e}^a_j,\mathbf{e}^b_t)}\le \sqrt{2}\,.
\end{equation}
If for these LHVMs we have the basic experimental fact $P(\mathbf{n},\mathbf{n})=-1$ for the Bell state, then Eq.~\eqref{eq:StringentRestrictions4PHVM} already shows the violation of CHSH inequality. This is the inherent incompatibility of LHVMs which have both the angular momentum conservation and the bilinear form of spin correlation. 

Note that the counterexample of Eq.~\eqref{eq:ResultofCounterexample} and the analysis of the inherent incompatibility both are based on Bell state, and not all spin states of collision products are exactly equal to the Bell state, so to some extent the LHVMs in collider experiments can evade these analyses. Therefore, we need a more rigorous approach to negate the feasibility of testing Bell's inequalities using scattering cross-section data.

\section{An LHVM with Exact Scattering Cross-Sections}
\label{sec:LHVMwithExactScatt}

To conclusively debunk the possibility of testing Bell inequalities at colliders, we construct a specific LHVM that can give rise to the scattering cross-sections fully consistent with quantum field theory. Note that while it is relatively straightforward to construct LHVMs that satisfy all differential scattering cross-sections, such models fall short of providing a rigorous proof that scattering cross-section data cannot be used to test Bell inequalities. As an addition, we will also consider the decoherence that may occur in the intermediate processes, such as that induced by measurements. For simplicity, we assume that $\mathcal{X}+\mathcal{Y}$ has only a single reaction channel, with the products being $\mathcal{A}+\mathcal{B}$, and the decay channels of the particles $\mathcal{A}$ and $\mathcal{B}$ are also unique. Once we understand the construction process of this specific LHVM, extending it to the general cases will not be difficult.

As stated in the introduction, we assume that the collider detectors are extremely powerful, capable of directly detecting the angular distributions of the $a_{1\sim3}b_{1\sim3}$ particles as well as the angular distributions of the $\mathcal{AB}$ particles. Although in many cases the lifetimes of $\mathcal{A}$ or $\mathcal{B}$ are extremely short, differing from the usual long-lived particles, $\mathcal{A}+\mathcal{B}$ ultimately correspond to some types of states, and their respective momenta can be defined. Therefore, in principle, we can directly measure the outgoing angles of $\mathcal{A}$ and $\mathcal{B}$ (which is not achievable with current technology). The reason for assuming such a powerful collider detector is just to demonstrate that even if some extremely advanced collider detection technology could be developed, as long as spin cannot be measured in an arbitrary direction, the test of Bell's inequalities cannot be conducted. With such a powerful collider, our LHVM must have the same scattering cross section as quantum field theory at all measurement levels.

To clarify the difficulties in constructing a general LHVM, we first consider a naive LHVM. Suppose we have obtained the expression for the differential scattering cross section of the process $\mathcal{XY}\to\mathcal{A}^s\mathcal{B}^{\bar{s}}$ through analytical calculation (at leading order for simplicity), which is a function of momentum, and then directly use this function to construct an LHVM so that the angular distribution of $\mathcal{A}^s+\mathcal{B}^{\bar{s}}$ is consistent with this differential scattering cross section. To be consistent with experimental detection, we also need this LHVM to correctly yield the angular distributions of the decay products of $\mathcal{A}^s$ and $\mathcal{B}^{\bar{s}}$. Unfortunately, such an LHVM cannot be consistent with QFT at all levels of the scattering cross section. Specifically, if we do not measure the intermediate states but measure the angular distributions of the $a_{1\sim3}b_{1\sim3}$ particles, this LHVM will yield the following scattering cross section:
\begin{equation}
\dv{\sigma}{\Pi_{a_{1\sim 3}}\dd\Pi_{b_{1\sim 3}}}\propto \sum_{s,\bar{s}}\abs{\mathcal{M}(\mathcal{XY}\to\mathcal{A}^s\mathcal{B}^{\bar{s}})}^2\abs{\mathcal{M}(\mathcal{A}^s\to a_{1\sim 3})}^2\abs{\mathcal{M}(\mathcal{B}^{\bar{s}}\to b_{1\sim 3})}^2\,.\label{eq:CrossSectionInNaiveLHVM}
\end{equation}
In quantum field theory, if using the narrow width approximation, this scattering cross section should be
\begin{equation}
\dv{\sigma}{\Pi_{a_{1\sim 3}}\dd\Pi_{b_{1\sim 3}}}\propto \abs{\sum_{s,\bar{s}}\mathcal{M}(\mathcal{XY}\to\mathcal{A}^s\mathcal{B}^{\bar{s}})\mathcal{M}(\mathcal{A}^s\to a_{1\sim 3})\mathcal{M}(\mathcal{B}^{\bar{s}}\to b_{1\sim 3})}^2\,.\label{eq:CrossSectionWithNWA}
\end{equation}
The main difference between Eq.~\eqref{eq:CrossSectionInNaiveLHVM} and Eq.~\eqref{eq:CrossSectionWithNWA} is the absence of interference effects in Eq.~\eqref{eq:CrossSectionInNaiveLHVM}, so this LHVM can easily be ruled out by the collider detector assumed here. To account for interference effects, we must construct a more powerful LHVM. Next, we will show how to construct the required LHVM.

\begin{figure}[ht]
\centering 
\includegraphics[width=0.7\textwidth]{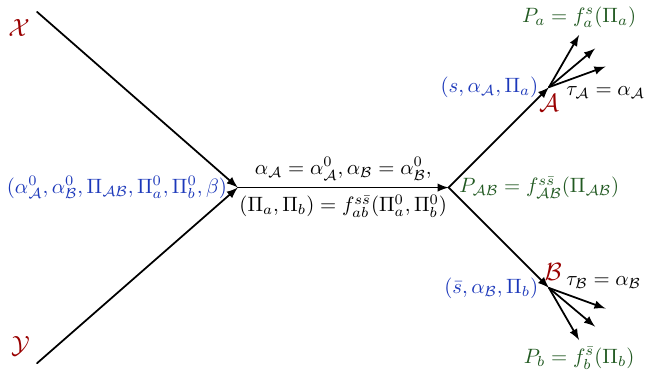}
\caption{\label{fig:counterexampleLHVM} The relationships between hidden variables, and the relationships between hidden variables and final state momenta. When decoherence occurs, the hidden variables will redistribute.}
\end{figure}

A schematic representation of this specific LHVM is shown in  Fig.~\ref{fig:counterexampleLHVM}. We assume that the decay of particle $\mathcal{A}^{s}$ involves a set of hidden variables $(\alpha_\mathcal{A},\Pi_a)$, where $\alpha_\mathcal{A}$ is a real parameter and the values of $\Pi_a$ have one-to-one correspondence to the points in the final state phase space of the decay of $\mathcal{A}$. If decoherence occurs, $\alpha_\mathcal{A}$ will follow an exponential distribution with mean $1/\Gamma^\mathcal{A}$, and $\Pi_a$ will follow a uniform distribution on the decay final state phase space; otherwise, $\alpha_{\mathcal{A}}$ and $\Pi_a$ are determined by the reaction at the previous level. Here $\alpha_\mathcal{A}$ determines the lifetime of particle $\mathcal{A}$, $\Delta t=\alpha_\mathcal{A}$, and $\Pi_a$ determines the four-momentum state of the decay products through a transformation $f_a^s$ on the decay final state space: $P_a=f_a^s(\Pi_a)$. According to quantum field theory, we know that
\begin{equation}
\dv{\Gamma^{\mathcal{A}^s}}{V_a}\propto \abs{\mathcal{M}(\mathcal{A}^s\to a_1a_2a_3)}^2\,,
\end{equation}
where $\dd V_a$ is a volume element on the final state phase space of the decay of $\mathcal{A}$. We just need to choose a $f_a^s$ such that
\begin{equation}
\frac{1}{\Gamma^{\mathcal{A}^s}}\dv{\Gamma^{\mathcal{A}^s}}{V_a}\abs{\det(\pdv{f_a^s}{\Pi_a})}=\frac{1}{V_a}
\end{equation}
to ensure that the distribution of decay products of particle $\mathcal{A}^s$ in the final state space satisfies the predictions of quantum field theory. According to a weaker version of Moser's theorem~\cite{moser1965volume}, such a $f_a^s$ always exists, and it is a function that transforms the uniform distribution on the decay final state space of $\mathcal{A}$ into a distribution proportional to $\abs{\mathcal{M}(\mathcal{A}^s\to a_1a_2a_3)}^2$. Similarly, we can construct hidden variables $(\alpha_\mathcal{B},\Pi_b)$ for particle $\mathcal{B}^{\bar{s}}$.

Next, we assume that the collision of $\mathcal{X}$ and $\mathcal{Y}$ involves a set of hidden variables $(\alpha_\mathcal{A}^0, \alpha_\mathcal{B}^0, \Pi_{\mathcal{AB}}, \Pi^0_{a}, \Pi^0_b, \beta)$, where $\alpha_\mathcal{A}^0$ and $\alpha_\mathcal{B}^0$ are real parameters, following exponential distributions with means $1/\Gamma^\mathcal{A}$ and $1/\Gamma^\mathcal{B}$, respectively. Under total momentum constraints, $\Pi_{\mathcal{AB}}$ corresponds to points on the final state phase space of $\mathcal{A}+\mathcal{B}$, following a uniform distribution. $\Pi^0_{a}$ and $\Pi^0_b$ correspond to points on the decay final state phase space of $\mathcal{A}$ and $\mathcal{B}$, respectively, and both follow uniform distributions on their respective phase spaces. $\beta$ is an additional set of parameters used to determine the spin of the collision products $\mathcal{A}$ and $\mathcal{B}$, for example, obtaining the probability of $\mathcal{A}^s+\mathcal{B}^{\bar{s}}$ as $\sigma^{s\bar{s}}/\sigma$. After the collision, for the newly produced particles the $\alpha$ parameters are equal to their corresponding $\alpha^0$, and the position of $\mathcal{A}$ and $\mathcal{B}$ in the final state phase space, $P_{\mathcal{AB}}$, is determined by $\Pi_{\mathcal{AB}}$ and a phase space transformation $f_{\mathcal{AB}}^{s\bar{s}}$: $P_{\mathcal{AB}}=f_{\mathcal{AB}}^{s\bar{s}}(\Pi_{\mathcal{AB}})$; the hidden variables $\Pi_a$ and $\Pi_b$ of $\mathcal{A}$ and $\mathcal{B}$ are determined by ($\Pi_a^0$, $\Pi_b^0$) and a mapping $f_{ab}^{s\bar{s}}$: $(\Pi_a,\Pi_b)=f_{ab}^{s\bar{s}}(\Pi_a^0,\Pi_b^0)$. It is easy to define the transformation $f^{s\bar{s}}_{\mathcal{AB}}$, it just needs to satisfy
\begin{equation}
\frac{1}{\sigma^{s\bar{s}}}\dv{\sigma^{s\bar{s}}}{V_{\mathcal{AB}}}\abs{\det(\pdv{f_{\mathcal{AB}}^{s\bar{s}}}{\Pi_{\mathcal{AB}}})}=\frac{1}{V_{\mathcal{AB}}}\,,
\end{equation}
where $\dd V_\mathcal{AB}$ is the volume element of the final state phase space of $\mathcal{A}+\mathcal{B}$. Next, we need to determine $f_{ab}^{s\bar{s}}$, which requires a more complex treatment because we must consider the interference effects. For this, we need to take the direct product of the decay final state phase space of $\mathcal{A}$ and the decay final state phase space of $\mathcal{B}$ to form a higher-dimensional compact manifold $M$, and $f_{ab}^{s\bar{s}}$ is a transformation on $M$. According to our settings for the hidden variables, after collision and particle decay, the distribution of final state particles is
\begin{eqnarray}
& & \int\dd V_{\mathcal{AB}} \sum_{s,\bar{s}}\Bigg\{ \frac{\sigma^{s\bar{s}}}{\sigma}\cdot\frac{1}{\sigma^{s\bar{s}}}\dv{\sigma^{s\bar{s}}}{V_{\mathcal{AB}}}\cdot\frac{\delta(h(P_{\mathcal{AB}},P_a,P_b))}{V_aV_b}\nonumber \\
&&\hspace{2.5cm} \times
\abs{\det(\pdv{f^{s\bar{s}}_{ab}}{(\Pi_a^0,\Pi_b^0)})}^{-1}\abs{\det(\pdv{f_a^s}{\Pi_a})}^{-1}\abs{\det(\pdv{f_b^{\bar{s}}}{\Pi_b})}^{-1}\Bigg\} \nonumber  \\
= &&  \sum_{s,\bar{s}}\frac1{\sigma}\dv{\sigma^{s\bar{s}}}{V_{\mathcal{AB}}}\cdot\frac{F(P_a,P_b)}{\Gamma^{\mathcal{A}^s}\Gamma^{\mathcal{B}^{\bar{s}}}}\dv{\Gamma^{\mathcal{A}^s}}{V_a}\dv{\Gamma^{\mathcal{B}^{\bar{s}}}}{V_b}\abs{\det(\pdv{f^{s\bar{s}}_{ab}}{(\Pi_a^0,\Pi_b^0)})}^{-1}\,,
\end{eqnarray}
where the function $h(P_{\mathcal{AB}},P_a,P_b)$ represents the constraint relationship between the phase space position of the decay particles and the phase space position of $\mathcal{A}$ and $\mathcal{B}$ particles, arising from energy-momentum conservation. $F(P_a,P_b)$ is a non-negative factor obtained after integrating out the Dirac delta function. Next, we select a set of $f_{ab}^{s\bar{s}}$ that satisfy
\begin{eqnarray}
\sum_{s,\bar{s}}\frac1{\sigma}\dv{\sigma^{s\bar{s}}}{V_{\mathcal{AB}}}\cdot\frac{F(P_a,P_b)}{\Gamma^{\mathcal{A}^s}\Gamma^{\mathcal{B}^{\bar{s}}}}\dv{\Gamma^{\mathcal{A}^s}}{V_a}\dv{\Gamma^{\mathcal{B}^{\bar{s}}}}{V_b}\abs{\det(\pdv{f^{s\bar{s}}_{ab}}{(\Pi_a^0,\Pi_b^0)})}^{-1} =\frac{1}{\sigma}\frac{\dd{}^2\sigma}{\dd V_a\dd V_b}\,.
\end{eqnarray}
Such $f_{ab}^{s\bar{s}}$ can take many forms, for example, we can choose $f_{ab}^{s\bar{s}}$ to be independent of $s$ and $\bar{s}$, which gives
\begin{equation}
\frac{\dd{}^2\sigma}{\dd V_a\dd V_b}\abs{\det(\pdv{f^{r\bar{r}}_{ab}}{(\Pi_a^0,\Pi_b^0)})}\\=\sum_{s,\bar{s}}\dv{\sigma^{s\bar{s}}}{V_{\mathcal{AB}}}\cdot\frac{F(P_a,P_b)}{\Gamma^{\mathcal{A}^s}\Gamma^{\mathcal{B}^{\bar{s}}}}\dv{\Gamma^{\mathcal{A}^s}}{V_a}\dv{\Gamma^{\mathcal{B}^{\bar{s}}}}{V_b}\,.
\end{equation}
Now the specific LHVM we need has been constructed.

This LHVM not only yields the correct final scattering cross-section but also provides the correct scattering cross-section for intermediate processes, and it can even reflect changes in the scattering cross-section after decoherence occurs. Specifically, whether the measured scattering cross section is $\mathcal{XY} \to \mathcal{AB}$, $\mathcal{XY} \to a_{1\sim3}b_{1\sim3}$, or even $\mathcal{XY} \to \mathcal{A} + b_{1\sim3}$, this LHVM can yield a spin-summed scattering cross section consistent with quantum field theory. Although our argument here is based on the assumption that $\mathcal{X} + \mathcal{Y}$ has only the single reaction channel $\mathcal{A} + \mathcal{B}$, it can be generalized to more common situations. However, this requires a clear definition of the quantum state $\mathcal{A} + \mathcal{B}$, because in quantum field theory $\mathcal{X} + \mathcal{Y}$ can directly transition to $a_{1\sim3} + b_{1\sim3}$ without becoming $\mathcal{A} + \mathcal{B}$. Once the quantum state of $\mathcal{A} + \mathcal{B}$ is well-defined, introducing additional hidden variables can handle the situation with multiple reaction channels. On the other hand, if the collider detector measures the momenta of particles in the intermediate process, the particle wavefunctions collapse, and the subsequent decays of the particles will no longer exhibit interference effects. Our LHVM will also correctly produce the angular distributions of the decay products, which could not achieved by the naive LHVM mentioned at the beginning of this section. Based on the LHVM constructed here, we assert that whether measuring the final scattering cross section, the intermediate process scattering cross section, or the angular distributions of products at multiple different levels during the reaction, none can be used to conduct a test of Bell's inequalities. 


Note that the LHVM constructed here is not necessarily a physical model in the real world. Instead, we merely use it as a counterexample for the test of Bell's inequalities. The possibility of constructing such a counterexample is entirely due to the fact that collider measurements involve momentum eigenstates. However, this does not imply that detectors like the Stern-Gerlach apparatus cannot be used for Bell tests, even though these devices ultimately measure particle positions or momenta. The essential difference lies in whether the spin degree of freedom is entangled with the position or momentum degrees of freedom. For instance, a pure state of silver atoms passing through a Stern-Gerlach apparatus would result in a quantum state becoming
\begin{equation}
c_1\ket{\uparrow}\otimes\ket{\text{position above}}+c_2\ket{\downarrow}\otimes\ket{\text{position below}}\,.
\end{equation}
Such entangled states are key to realizing spin measurements. This analysis is based on standard quantum mechanics. In a general theory encompassing LHVMs, what the Stern-Gerlach apparatus measures is not crucial, this quantity can be completely separated from physical quantities such as spin, momentum, etc. It is merely the probabilistic outcome (or deterministic, considering hidden variables) of the interaction between the input state and the measuring device. For example, we assume that silver atoms exiting the Stern-Gerlach apparatus will hit a screen, thus the screen is the device that realizes the measurement, causing the wave function to collapse. Therefore, the physical quantity that silver atoms are ultimately measured by is position rather than spin. However, this point is not crucial, what is important is that we can obtain a quantity based on the final position of the silver atoms, which can be either $+1$ or $-1$, regardless of whether this quantity can be called spin. By performing multiple measurements on particle pairs using the Stern-Gerlach apparatus, we can define a statistical quantity $P(\mathbf{n}^a,\mathbf{n}^b)$, on which basis the Bell inequality can be derived by the standard process. Without doubt, this Bell inequality is consistent with the original Bell inequality, even though it originates from measurements of the position of silver atoms. Hence, it is insufficient to assert that Bell's inequalities cannot be tested merely because collider detectors measure momentum eigenstates~\cite{Abel:1992kz}. The rigorous argument comes from the LHVM we construct here: this LHVM indicates that there is no Bell-type inequality based on scattering cross-sections that can distinguish between standard quantum field theory and local hidden variable theories. The so-called test of Bell's inequalities on colliders merely applies the standard quantum mechanical a priori conclusions about spin measurements, and using such conclusions to rule out LHVMs is logically inconsistent. This loophole, differing fundamentally from other loopholes in Bell tests (such as the locality loophole~\cite{Severi:2021cnj} and the fair sampling loophole~\cite{Garg:1986wd}), originates entirely from logic and can be addressed by adding spin measurement devices on colliders, albeit measuring the spin of high-energy scattering states faces exceptional challenges.

\section{Testable LHVMs at Collider}
\label{sec:TestableLHVMs}

Based on our argument, to test the Bell inequalities at high energy scales, we must directly measure the spin of the scattering particles. The indirect testing results obtained using the scattering cross-sections from colliders may indicate the direction for future experiments, such as how to achieve the maximum violation of Bell inequalities~\cite{Cheng:2023qmz}. However, if an LHVM satisfies Eq.~\eqref{eq:EqforCij} where $C_{ij}$ is the spin correlation in the direction of the basis vectors, then such an approach can indeed test the Bell inequalities for it, and we denote the set of such LHVMs as $L_\sigma$. Furthermore, we denote the set of all LHVMs as $L_{\rm total}$, and the set of LHVMs where the spin correlation is in a bilinear form as $L_C$. Then we have
\begin{equation}
L_\sigma\subsetneq L_C\subsetneq L_{\rm total}\,.
\end{equation}

Next, we present a class of LHVMs to demonstrate that $L_\sigma$ is not an empty set. We assume that $\mathcal{A}$ and $\mathcal{B}$ have definite spins from the moment of their creation, with spin orientations denoted as $\mathbf{u}^a$ and $\mathbf{u}^b$, respectively.
The distribution of $\mathbf{u}^a$ and $\mathbf{u}^b$ is given by $D(\mathbf{u}^a,\mathbf{u}^b)$, which can be considered as being induced by hidden variables.

We can further introduce additional hidden variables to describe the subsequent behavior of $\mathcal{A}$ and $\mathcal{B}$, though we may not explicitly represent them. We only require that the statistical behavior caused by these hidden variables is consistent with the quantum statistical behaviors of $\mathcal{A}$ and $\mathcal{B}$ respectively. For example, when we measure the spin of particle state $\ket{\mathbf{u}^{a}}$ along $\mathbf{n}^a$, the probability of obtaining $+1$ should be consistent with quantum mechanics. Such hidden variables exist, for example, the model mentioned earlier with $\mathbf{m}^{\mathcal{A}}$ and $\mathbf{m}^{\mathcal{B}}$ as hidden variables (see Eq.~\eqref{eq:BellLHVM}). Therefore, when the spin orientations of $\mathcal{A}$ and $\mathcal{B}$ are $\mathbf{u}^a$ and $\mathbf{u}^b$ respectively, the spin correlation measured in the directions $\mathbf{n}^a$ and $\mathbf{n}^b$ is $(\mathbf{n}^a\cdot\mathbf{u}^a)(\mathbf{u}^b\cdot\mathbf{n}^b)$, and thus the overall spin correlation is
\begin{equation}
P(\mathbf{n}^a,\mathbf{n}^b)=\int\dd{\mathbf{u}^a}\dd{\mathbf{u}^b}\,(\mathbf{n}^a\cdot\mathbf{u}^a)(\mathbf{u}^b\cdot\mathbf{n}^b)D(\mathbf{u}^a,\mathbf{u}^b)\,.
\end{equation}
We can see that in this case, $P(\mathbf{n}^a,\mathbf{n}^b)$ has a bilinear form, and the spin correlation coefficients are 
\begin{equation}
C_{ij}=\int\dd{\mathbf{u}^a}\dd{\mathbf{u}^b}D(\mathbf{u}^a,\mathbf{u}^b)u^a_iu^b_j\,.
\end{equation}
Furthermore, since we have assumed that the statistical properties of individual particles in these LHVMs are consistent with quantum mechanics, therefore the differential decay width of $\mathcal{A}$ with spin orientation $\mathbf{u}^a$ is
\begin{equation}\label{eq:DifferentialDecayWidthA}
\begin{aligned}
\dv{\Gamma^{\mathcal{A}}}{V_a} &\propto \braket{a_{1\sim 3}}{\mathbf{u}^a}\braket{\mathbf{u}^a}{a_{1\sim 3}}\\
&= \sum_{r,s}\frac12(\delta_{rs}+u^a_i\sigma_{i,rs})\Gamma^{\mathcal{A}}_{sr}\,,
\end{aligned}
\end{equation}
where $\Gamma_{sr}^{\mathcal{A}}$ is consistent with the decay spin density matrix of quantum field theories. If we define
\begin{eqnarray}
    B_i^{\mathcal{A}} &=& \int\dd{\mathbf{u}^a}\dd{\mathbf{u}^b}D(\mathbf{u}^a,\mathbf{u}^b)u^a_i\,,\\
    B_i^{\mathcal{B}} &=& \int\dd{\mathbf{u}^a}\dd{\mathbf{u}^b}D(\mathbf{u}^a,\mathbf{u}^b)u^b_i\,,
\end{eqnarray}
we will obtain
\begin{eqnarray}\label{eq:NewDifferentialCrosection}
\begin{aligned}
\frac{1}{\sigma}\frac{\dd{}^4\sigma}{\dd{}^2\Omega^\mathcal{A}\dd{}^2\Omega^\mathcal{B}} &= \frac{1}{(4\pi)^2}\int\dd{\mathbf{u}^a}\dd{\mathbf{u}^b}D(\mathbf{u}^a,\mathbf{u}^b)(1+\kappa^{\mathcal{A}}u^a_i\Omega_i^{\mathcal{A}})(1+\kappa^{\mathcal{B}}u^b_j\Omega_j^{\mathcal{B}})\\
&= \frac{1}{(4\pi)^2}\big(1+\kappa^\mathcal{A}B_i^\mathcal{A}\Omega_i^\mathcal{A}+\kappa^\mathcal{B}B_i^\mathcal{B}\Omega_i^\mathcal{B} +\kappa^\mathcal{A}\kappa^\mathcal{B}\Omega_i^\mathcal{A}C_{ij}\Omega_j^\mathcal{B}\big)\,.
\end{aligned}
\end{eqnarray}
It has the same form as Eq.~\eqref{eq:EqforCij}. Thus we can use the standard method to extract $C_{ij}$ and use it to test Bell inequalities. This type of LHVMs attempt to interpret quantum correlation as classical correlation, which is very intuitive in classical physics. All assumptions except the classical correlation are based on quantum mechanics, making it a very attractive class of LHVMs.

\begin{figure}[ht]
\centering 
\includegraphics[width=0.4\textwidth]{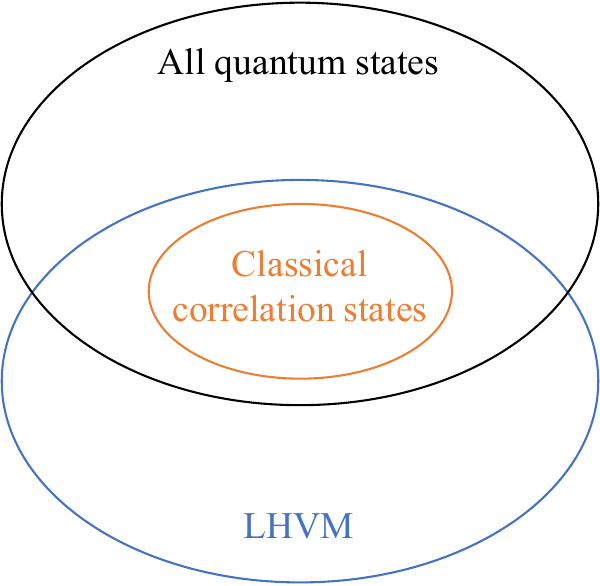}
\caption{\label{fig:KindsOfQuantumState} All quantum states with classical correlations can be described by LHVMs and can be tested using the Peres-Horodecki criterion. The states that can be described by LHVMs include not only those with classical correlations but also those with non-classical correlations, and they satisfy Bell's inequalities.}
\end{figure}

Physical states in this kind of LHVMs are convex sum of infinite separable states, thus the Peres-Horodecki separability criterion can also be used to rule out this type of LHVMs~\cite{peres1996separability,Horodecki19961}. Note that both spin correlation coefficients $C_{ij}$ and spin polarization $B^{\mathcal{A}}_i,B_i^{\mathcal{B}}$ of final states are required to determinate the separability of physical states~\cite{Afik_2021, Afik_2023}. Since there are inseparable states that can be described by LHVMs according to Werner's example~\cite{werner1989quantum}, the Peres-Horodecki inseparability is expected to achieve a higher confidence level than violation of Bell inequalities for this kind of LHVMs. Fig.~\ref{fig:KindsOfQuantumState} illustrates the reasons for this in the form of a Venn diagram.

On the other hand, once a two-qubit quantum state (whether separable or not) can be described by an LHVM, then this LHVM can be extended to a model belonging to $L_\sigma$ by adding special scattering and decay mechanics. The feasibility of this approach stems from the commutativity of spin operators and momentum operators. Therefore, testing Bell inequalities via scattering cross-section data could rule out this kind of LHVMs describing inseparable states contained in $L_\sigma$ while the Peres-Horodecki criterion becomes invalid.

\section{Conclusions}
\label{sec:conclusions}

We analyzed the incompatibilities arising from the bilinearity of the spin correlation function, and constructed an LHVM whose scattering cross-section is fully consistent with standard quantum theory, thereby demonstrating that in general Bell inequalities cannot be tested using collider scattering cross-section data unless we can directly measure the spins of the outgoing particles or introduce additional assumptions. The LHVM we constructed is very robust. Whether the collider detector measures the scattering cross section of the final products, the scattering cross section of the intermediate particles, or even measures the scattering cross section of the intermediate particles followed by the angular distributions of their decay products (which would result in the disappearance of interference effects), our LHVM will yield predictions consistent with quantum field theory. Therefore, even if the collider detector is powerful enough to provide scattering cross-section data at all levels, it still cannot be used to test Bell's inequalities. Meanwhile, we found that if an LHVM employs classical spin correlations as a surrogate for quantum spin correlations while preserving other statistical properties of quantum theory, its Bell inequalities can be tested via scattering cross-section data. 

\section*{Acknowledgments}
\phantomsection 
\addcontentsline{toc}{section}{Acknowledgments}
This work was supported by the National Natural Science Foundation of China under grant numbers 11821505, 12075300 and 12335005, the Peng-Huan-Wu Theoretical Physics Innovation Center under grant number 12047503, and the Key Research Program of the Chinese Academy of Sciences under grant number XDPB15.

\phantomsection 
\addcontentsline{toc}{section}{References}
\bibliographystyle{JHEP}
\bibliography{bibliography}

\end{document}